\documentstyle[eqsecnum,aps,preprint,epsfig]{revtex}
\begin{document}


\draft
\tightenlines
\title{Aspects of high density effective theory in QCD}

\author{Deog Ki Hong\footnote{On leave from
Department of Physics, Pusan National University,
Pusan 609-735, Korea. \\
E-mail address: {\tt dkhong@hyowon.cc.pusan.ac.kr}
}}

\address{
Lyman Laboratory of Physics,
Harvard University, Cambridge, MA 02138
\protect\\   and \protect\\
Physics Department, Boston University,
Boston, MA 02215, USA}


\maketitle

\begin{abstract}
We study an effective theory of QCD at high density
in detail, including the finite temperature effects and the
leading order correction in $1/\mu$ expansion.
We investigate the Cooper pair gap equation and find that the
color-flavor locking phase is energetically preferred at high
density.
We also find the color-superconducting phase transition occurs
in dense quark matter when the chemical
potential is larger than $250\pm 100~{\rm MeV}$ and the temperature
is lower than $0.57$ times the Cooper pair gap in the leading order
in the hard-dense-loop approximation.
The quark-neutrino four-Fermi coupling and the quark-axion coupling
receive  significant corrections in dense quark matter.
\end{abstract}
\vskip 0.2in

\section{Introduction}
\label{sec:label1}

Effective field theory is indispensable to describe physical
processes at energy lower than the characteristic scale of theory,
especially when the microscopic physics is too complicated to be used
as in the case of chiral perturbation theory~\cite{weinberg} or just
unknown as in the standard model. Since effective field
theory has been very useful and successful in explaining the low
energy data, it has now become an everyday language in
physics~\cite{kaplan,polchinski}.

Color superconductivity in cold and dense quark
matter~\cite{bailinlove} has been studied quite intensively in recent
years by using effective field theory
methods~\cite{shuryak,wilczek,wilczek2,nick,rho}. Most important
ingredient in those studies was the existence of effective four-quark
interactions that mimic the four-fermion interaction generated by
phonon exchange in BCS superconductivity, which are assumed to be
generated in the effective Lagrangian either by exchange of massive
gluons~\cite{nick} or by strongly coupled
instantons~\cite{shuryak,wilczek}.
For low density quark matter, where strong interaction is no
longer weak, such four-Fermi interactions may
describe QCD at low density as model interactions, thus leading to
color superconducting gap of QCD scale ($\sim\Lambda_{\rm QCD}$).
But, it was recently argued quite convincingly by Son~\cite{son}
that at least for high density quark matter, where perturbative
QCD is applicable, magnetic gluons are not Debye-screened but only
dynamically screened or Landau damped.  Such long-range magnetic
gluon interaction is more important and leads to a bigger Cooper-pair
gap, $\Delta\sim\mu g_s^{-5}\exp(-3\pi^2/\sqrt{2}g_s)$,
than the usual BCS gap,
which is subsequently confirmed by the Schwinger-Dyson analysis for
the Cooper-pair gap~\cite{hong98,hmsw,sw99,pr99}.

Recently, an effective field theory of QCD at high
density~\cite{hong98}
is derived systematically in the power expansion of $1/
\mu$ by integrating out the
anti-quarks which are decoupled at low energy in dense quark
matter. In the effective theory,
four-quark operators are generated at the one-loop matching and
the electric gluons are screened due to
quarks in the Fermi sea while the magnetic gluons are not at least
in perturbation~\cite{son,pisarski,kapusta88,manuel,nair}.
At scales below the electric
screening mass, the relevant interactions for quarks are the coupling
with magnetic gluons and the four-quark interaction with opposite
momenta. Both interactions for quarks in the color anti-triplet
channel with opposite momenta are shown to lead to color
superconductivity~\cite{son,hong98}.

According to the quark-hadron phase continuity in QCD, conjectured
recently by Sch\"afer and Wilczek~\cite{schafer},
the confining phase of nuclear matter at low density is complementary
to the Higgs phase quark matter at high density.
We support this conjecture by showing that in the effective theory,
which is equivalent to QCD in the asymptotic density,
the color-flavor locking diquark condensate is energetically
more preferred.
Furthermore, in the effective theory, the higher order corrections
are systematically calculable and one can easily estimate quantities
like the critical density, which are determined by the sub-leading
operators in $1/\mu$ expansion.

In this article, we study the effective lagrangian of QCD at high
density derived in~\cite{hong98} in detail,
including the higher order corrections in $1/\mu$
expansion and the finite temperature effects.
We then analyze the Cooper-pair gap equation more rigorously and
calculate the gap, the critical temperature, and the critical density.
As applications, we examine the effect of marginal
four-quark operators on the neutrino-quark four-Fermi coupling
and the quark-axion coupling in super-dense quark matter.


\section{high density effective theory}
\label{sec:label2}


The idea of effective field theory is quite simple and
its construction can be performed in a systematic way.
First, one identifies or guesses the right degrees of freedom to
describe the low energy processes.  Then, one integrates
out the irrelevant degrees of freedom, which results in nonlocal
interactions. By expanding the nonlocal interactions in powers of
momentum at low energy, one derives a (Wilsonian) effective
Lagrangian, which is usually done in practice by matching
all the one light-particle irreducible amplitudes in the full theory
with those amplitudes in the effective theory in loop expansion.
Since the higher dimensional operators have smaller effects at low
energy, the effective theories provide a very useful approach to
describe low energy physics.

A system of degenerate quarks with a fixed baryon number is described
by the QCD Lagrangian density with a chemical potential $\mu$,
\begin{equation}
{\cal L}_{\rm QCD}=\bar\psi
i\!\not\!D\psi-{1\over4}F_{\mu\nu}^aF^{a\mu\nu}+\mu
\bar\psi\gamma_0\psi,
\label{lag}
\end{equation}
where the covariant derivative
$D_{\mu}=\partial_{\mu}+ig_sA_{\mu}^aT^a$ and
we neglect the mass of quarks for simplicity.
The chemical potential is introduced as a Lagrangian
multiplier to constrain the system to have a fixed baryon number,
$N_B$. At zero temperature, all the states up to the Fermi surface are
filled;
\begin{equation}
{N_B\over V}=2N_f\int{{\rm d}^3p\over (2\pi)^3}\theta(p_F-\left|\vec
p\right|),
\end{equation}
where $V$ is the volume, $N_f$ the number of quark flavors,
and $\vec p_F$ is the Fermi momentum.
The chemical potential $\mu$, which defines the energy at the Fermi
surface, provides another scale for QCD.
Here we consider a super-dense quark matter where quarks are
very dense so that the average inter-quark distance is much smaller
than  the characteristic length of QCD, $1/\mu\ll1/\Lambda_{\rm QCD}$.
For high chemical potential, the pair creation of a particle and
an anti-particle is suppressed at low energy because of the energy
gap $\mu$, provided by the Fermi sea. Therefore, the relevant degrees
of freedom at low energy will be just particles and holes,
excited near the Fermi surface, together with gluons.


We will derive a low energy effective Lagrangian for cold and dense
quark matter by integrating out the high
frequency or irrelevant modes, {\it \'a la} Wilson,
as we run QCD from high energy $E>\mu$ to low
energy $E\sim \Lambda_{\rm QCD}$ . Since QCD is
asymptotically free, quarks will interact weakly at energy
$E>\mu~(\gg\Lambda_{\rm QCD})$ and their spectrum will be
approximately given by the
free Hamiltonian:
\begin{equation}
\left(\vec\alpha\cdot \vec p-\mu\right)\psi_{\pm}=E_{\pm}\psi_{\pm},
\label{energy}
\end{equation}
where $\vec\alpha=\gamma_0\vec\gamma$ and $\psi_{\pm}$ denote the
energy eigenfunctions with eigenvalues $E_{\pm}=-\mu\pm \left|\vec
p\right|$, satisfying $\vec \alpha\cdot\vec
p\psi_{\pm}=\pm\left|\vec p\right|\psi_{\pm}$, respectively. At low
energy $E<\mu$, the states $\psi_+$ near the Fermi surface, $|\vec
p|\sim\mu$, are easily excited, while $\psi_-$, which correspond
to the states in the Dirac sea, are completely decoupled due to
the presence of the energy gap $\mu$ provided by the Fermi sea.
Therefore the right degrees of freedom below $\mu$ consist of
gluons and $\psi_+$ only.

If $\mu=0$, integrating out the high frequency modes in QCD just
renormalizes the coupling constants without generating any new
interactions as it should be for a renormalizable theory like QCD.
However, when $\mu\ne0$, the energy spectra of particles and
anti-particles are asymmetric. Anti-particles ($\bar\psi_-$), which
correspond to the absence of states ($\psi_-$) in the Dirac sea, have
energy larger than $\mu$ and will be decoupled at energy scale lower
than $\mu$. Therefore, the minial quark-gluon coupling that connects
$\psi_-$ with $\bar\psi_+$ (or vice versa),  has to be removed at low
energy, generating a new interaction among $\psi_+$.  We will see
later that integrating out the modes whose frequency is higher than
$\mu$ results in an effective Lagrangian which is
drastically different from QCD.

We now integrate out the antiparticles ($\bar\psi_-$) or the states
($\psi_-$) in the Dirac sea to derive the low energy effective
Lagrangian.
We first decompose the quark momentum into the Fermi momentum and a
residual momentum as
\begin{equation}
p_{\mu}=\mu v_{\mu}+l_{\mu},
\label{decompo}
\end{equation}
where the residual momentum $|l_{\mu}|<\mu$ and $v^{\mu}=(0,\vec
v_F)$ with Fermi velocity $\vec v_F$.  The magnitude of Fermi
velocity, $\left|\vec v_F\right|\equiv p_F/\sqrt{p_F^2+m^2}$,
will be taken to be unity, for we assume the quark mass $m=0$.

Since we will integrate out the high frequency modes such that we will
be left with excitations only near the Fermi surface at low energy, it
is convenient to decompose the quark field by the Fermi
velocity~\footnote{For a given quark momentum, the corresponding Fermi
velocity is determined up to reparametrization; $\vec v_F\to \vec
v_F+\delta \vec l_{\perp}/\mu$ and $\vec l\to \vec l-\delta \vec l$,
where $\delta \vec l_{\perp}$ is a residual momentum perpendicular
to the Fermi velocity.
As in the heavy quark effective theory~\cite{manohar92},
the renormalization of
higher-order operators are restricted due to this
reparametrization invariance.}
\begin{equation}
\psi(x)=\sum_{\vec v_F} e^{i\mu\vec v_F\cdot \vec x}
\psi(\vec v_F,x),
\label{decomp}
\end{equation}
where
\begin{equation}
\psi(\vec v_F,x)=\int_{\left|l_{\mu}\right|<\mu} {{\rm
d}^4l\over (2\pi)^4}\psi(\vec v_F,l)e^{-il\cdot x}
\end{equation}
carries the residual momentum $l_{\mu}$.
We introduce projection operators $P_{\pm}=(1\pm\vec
\alpha\cdot\vec v_F)/2$ and define
\begin{equation}
\psi_{\pm}(\vec v_F,x)\equiv P_{\pm}\psi(\vec v_F,x).
\end{equation}
In the limit $l/\mu\to0$, $\psi_-(\vec v_F,x)$ corresponds to the
states in the Dirac sea and $\psi_+(\vec v_F,x)$ to the states above
the Dirac sea.

In terms of the new fields, the QCD Lagrangian for quarks at high
density becomes
\begin{eqnarray}
{\cal L}_{\rm quark}&=&\bar\psi(x)
\left(i\!\not\!D+\mu\gamma^0\right)\psi(x)\nonumber\\
&=&\sum_{\vec v_F}\left[\bar\psi_+(\vec
v_F,x)i\gamma^0V\cdot D\psi_+(\vec v_F,x)
+\bar\psi_-(\vec v_F,x)\gamma^0
\left(2\mu+i{\bar V}\cdot D\right)\psi_-
(\vec v_F,x)\right.
\label{qlag} \\
& & \left.
+\bar\psi_-(\vec v_F,x)i\gamma^{\mu}_{\perp}D_{\mu}\psi_+(\vec v_F,x)
+\bar\psi_+(\vec v_F,x)i\gamma^{\mu}_{\perp}D_{\mu}\psi_-(\vec v_F,x)
\right],
\nonumber
\end{eqnarray}
where $V^{\mu}=(1,\vec v_F)$, ${\bar V}^{\mu}=(1,-\vec v_F)$,
$\gamma^{\mu}_{\perp}=\gamma^{\mu}-\gamma^{\mu}_{\parallel}$
with $\gamma^{\mu}_{\parallel}=(\gamma^0,\vec v_F\vec v_F\cdot
\vec\gamma)$, and
we have used
\begin{eqnarray}
\bar\psi_+(\vec v_F,x)\gamma^{\mu}\psi_+(\vec v_F,x)&=&\bar\psi_+
P_-\gamma^{\mu}P_+\psi_+=V^{\mu}\bar\psi_+
\gamma^0\psi_+ \nonumber\\
\bar\psi_-(\vec v_F,x)\gamma^{\mu}\psi_-(\vec v_F,x)&=&\bar\psi_-
P_+\gamma^{\mu}P_-\psi_-={\bar V}^{\mu}\bar\psi_-
\gamma^0\psi_-
\\
\bar\psi_+(\vec v_F,x)\gamma^{\mu}\psi_-(\vec v_F,x)
&=&\bar\psi_+P_-\gamma^{\mu}P_-\psi_-=
\bar\psi_+\gamma^{\mu}_{\perp}\psi_-
\nonumber\\
\bar\psi_-(\vec v_F,x)\gamma^{\mu}\psi_+(\vec v_F,x)
&=&\bar\psi_-P_+\gamma^{\mu}P_-\psi_+=
\bar\psi_-\gamma^{\mu}_{\perp}\psi_+.
\nonumber
\end{eqnarray}
In Eq.~(\ref{qlag}) the incoming quark and the outgoing
quark have the same Fermi velocity $\vec v_F$,
since, at low energy, the momentum carried away by gluons can be
compensated by the residual momentum of the quarks
to conserve the momentum, without changing the Fermi velocity.

As we can see from Eq.~(\ref{qlag}), the propagation of quarks
at high density is 1+1 dimensional if $l^i_{\perp}/\mu$ is
negligible. This dimensional reduction at high density
can be also seen if we decompose the quark propagator
as following:
\begin{eqnarray}
iS_F(p)&=&{i\over (1+i\epsilon)p^0\gamma^0-\vec p\cdot \vec \gamma+
\mu\gamma^0}\nonumber\\
&=&{\not\! V\over 2}{i\over l\cdot V+i\epsilon
l_0}+{\not\!\bar V\over2}{i\over 2\mu+l\cdot\bar V+i\epsilon l_0}
+O(l^i_{\perp}/\mu),
\end{eqnarray}
where quark momentum  $p^{\mu}=\mu v^{\mu}+l^{\mu}$
and the $i\epsilon$ prescription is chosen such that
not only the anti-particles but also the holes correspond to the
negative energy states moving backward in time.
Note also that, if $l_{\perp}^i/\mu\to0$,  $1/2\!\not\!
V\gamma^0=1/2(1+\vec\alpha\cdot\vec v_F)$ is the projection operator
that projects out particles and holes ($\psi_+$), while $1/2\!\not\!
\bar V\gamma^0=1/2(1-\vec\alpha\cdot\vec v_F)$
projects out the states in the Dirac sea ($\psi_-$).
Since the excitation of quarks along the perpendicular direction to
the quark velocity does not cost any energy in the leading order,
the perpendicular momentum is nothing but an index that labels
the degeneracy in quark just as the degeneracy in the Landau level
under external constant magnetic field is labeled by the electron
momentum perpendicular to the external magnetic field.

At tree-level, integrating out the irrelevant modes $\psi_-(\vec
v_F,x)$ is tantamount to eliminating $\psi_-(\vec v_F,x)$ by
the equations of motion, given as
\begin{equation}
\psi_-(\vec v_F,x)=-{i\gamma^0\over 2\mu
+i{\bar D}_{\parallel}}\not\!D_{\perp}\psi_+(\vec v_F,x)=-
{i\gamma^0\over
2\mu}\sum_{n=0}^{\infty}\left(-{i{\bar D}_{\parallel}\over
2\mu}\right)^n\not\!D_{\perp}\psi_+(\vec v_F,x),
\label{eom}
\end{equation}
where ${\bar D}_{\parallel}={\bar V}^{\mu}D_{\mu}$
and $\not\!D_{\perp}=\gamma^{\mu}_{\perp}D_{\mu}$.
Plugging Eq.~(\ref{eom}) into the Lagrangian for quarks
Eq.~(\ref{qlag}),
we obtain the tree-level effective Lagrangian of
QCD at high density,
\begin{equation}
{\cal L}^0_{eff}=-{1\over4}F_{\mu\nu}^aF^{a\mu\nu}
+\sum_{\vec v_F}\left[\psi_+^{\dagger}(\vec
v_F,x)iV\cdot D\psi_+(\vec v_F,x)-\psi_+^{\dagger}
{(\not\!D_{\perp})^2\over 2\mu}
\sum_{n=0}^{\infty}\left(-{i{\bar D}_{\parallel}\over 2\mu}\right)^n
\psi_+(\vec v_F,x)\right].
\label{tree}
\end{equation}


\section{One-loop matching and a four-Fermi operator}
\label{sec:label3}

As we have seen in the previous section,
integrating out the
irrelevant modes generates a new coupling between quarks and gluons.
This can be also seen diagramatically as shown in Fig.~1:
In the leading order, the propagator of fast modes $\psi_-$ is
replaced by a constant matrix
\begin{equation}
{i\gamma^0\over 2\mu+i{\bar D}_{\parallel}}={i\gamma^0\over
2\mu}+O(l/\mu)
\end{equation}
and the exchange of $\psi_-$ generates a new interaction given as,
using $P_-\gamma^{\mu}P_-=P_-\gamma^{\mu}_{\perp}$,
\begin{equation}
{\cal L}^0_{\rm eff}\ni-{g^2\over2\mu}\sum_{\vec
v_F}\bar\psi_+\!\not\! A_{\perp}\gamma_0\!\not\!
A_{\perp}\psi_++\cdots,
\end{equation}
where the ellipsis
denotes terms containing more powers of gluons and derivatives.
By writing the interaction in a gauge-invariant fashion, we recover
the interaction terms in Eq.~(\ref{tree}).

The effect of integrating out fast modes not only generates
new interactions at low energy but also gives rise to quantum
corrections to the tree-level couplings which can be calculated
explicitly by matching the loop amplitudes.  Usually the new
interactions arise at tree-level upon integrating out the fast modes,
but sometimes they may arise at quantum level, especially when they
are marginal, as in the low energy effective theory
of QED under external (strong) magnetic field~\cite{hong97}.

As fermions under external fields~\cite{hong97,hong96,gusynin98},
in the presence of a Fermi sea~\cite{polchinski}
the scaling dimension of fields at low
energy changes due to the change in the particle spectrum; the
scaling dimension of quarks becomes $-1/2$ instead of $-3/2$, the
canonical scaling dimension of fermions in 3+1 dimensions.
The scaling dimension of fields is determined by
the kinetic term in the action, which is for quarks in dense quark
matter given in the high density limit as
\begin{equation}
S_0=\sum_{\vec v_F}\int {{\rm d}^4l\over(2\pi)^4}
\left[\psi_+^{\dagger}(\vec v_F,l) \left(l_0-v_F
l_{\parallel}\right)\psi_+(\vec
v_F,l)+\psi^{\dagger}_-(\vec v_F,l)\left(2\mu+l\cdot\bar V
\right)\psi_-(\vec v_F,l)\right],
\end{equation}
where $l_{\parallel}\equiv\vec v_F\cdot\vec l$ is the residual
momentum parallel to the Fermi velocity.
Under a scale transformation $l_0\to sl_0$ and
$l_{\parallel}\to sl_{\parallel}$ with $s<1$, the quark
fields transform as $\psi_+(\vec v_F,l)\to s^{-3/2}
\psi_+(\vec v_F,l)$ and $\psi_-(\vec v_F,l)\to s^{-1}
\psi_-(\vec v_F,l)$.
As we scale the momentum toward the Fermi surface or we
decrease $s\to0$, $\psi_+$ scales like free massless fermions
in (1+1)-dimensions, while $\psi_-$ scales like heavy fermions
in (1+1)-dimensions and becomes irrelevant at low energy.
Since four-Fermi interactions are
marginal  for (1+1)-dimensional (light) fermions,
we expect that a marginal
four-quark  operator will arise in the high density effective theory
of QCD  by quantum effects if it is absent at the tree-level. To see
this, we consider one-loop matching of a four-quark amplitude
for scattering of quarks with opposite Fermi velocities, shown in
Fig.~2, because only when the incoming quarks have opposite Fermi
momenta the four-quark operators are marginal as we scale toward the
Fermi surface~\cite{polchinski}. Since the effective theory amplitude
is ultra-violet (UV) divergent while the amplitude in full QCD is UV
finite by the power counting,  we need a UV counter term
for the one-loop matching in the
effective theory, which is nothing but a four-quark operator.  The
one-loop four-quark amplitude of our interest in the effective theory
is, reverting the notation $\psi$ for $\psi_+$ henceforth,
\begin{eqnarray}
A_4^{\rm eff}&=&{1\over2}\left(-{ig_s^2\over 2\mu}\right)^2
\sum_{\vec v_F,\vec v_F^{\prime}}
\left<\vec v_3,l_3;\vec
v_4,l_4\right| \int_{x,y}
\bar\psi\!\not\! A_{\perp}\gamma^0\!\not\! A_{\perp}\psi(\vec v_F,x)
\cdot\bar\psi\!\not\! A_{\perp}\gamma^0\!\not\! A_{\perp}\psi(\vec
v_F^{\prime},y) \left|\vec v_1,l_1;\vec v_2,l_2\right>
\nonumber\\
&=&-{g_s^4\over4\mu^2}(2\pi)^4\delta^4(\sum_i\! {}^{\prime}p_i)
\bar\psi(p_3)\gamma^{\mu}_{\perp}\gamma^0\gamma^{\nu}_{\perp}T^aT^b
\psi(p_1)\cdot\bar\psi(p_4)\gamma^{\rho}_{\perp}\gamma^0
\gamma^{\sigma}_{\perp}
T^cT^d\psi(p_2)\label{amp}
\\
&\times&\left(\delta^{ac}\delta^{bd}g_{\mu\rho}
g_{\nu\sigma}+\delta^{ad}\delta^{bc}g_{\mu\sigma}g_{\nu\rho}\right)
\cdot {-i\over
16\pi^2}\left[{1\over\epsilon}-\gamma-2-\ln\left({-l^2-i\epsilon\over
4\pi \Lambda^2}\right)\right],
\nonumber
\end{eqnarray}
where $\vec v_i$'s are the Fermi velocities of incoming and outgoing
particles with $\vec v_1=-\vec v_2$,
$p_i=\mu\vec v_i+l_i$, $\gamma$ the Euler number,
$\sum^{\prime}p_i\equiv p_3+p_4-p_1-p_2$, $l=l_3-l_1$,
and $\Lambda$ is the renormalization point.

In QCD, the quark-quark scattering amplitude is
\begin{eqnarray}
A_4^{\rm QCD}&=& {g_s^4\over 4!}
\left<p_3,p_4\right|\left[\sum_{\vec v_F}
\int_x\bar\psi(\vec v_F,x)\not\!A_{\perp}(x)\psi(\vec v_F,x)
\right]^4\left|p_1,p_2\right>\nonumber\\
&=&g_s^4(2\pi)^4\delta^4(\sum_i{}^{\prime}p_i)
\psi^{\dagger}(p_3)\gamma^{\mu}_{\perp}\gamma^{\nu}_{\perp}T^aT^b
\psi(p_1)
\psi^{\dagger}(p_4)\!\left[I_1\gamma_{\mu}\gamma_{\nu}
T^aT^b+I_2\gamma_{\nu}\gamma_{\mu}T^bT^a
\right]\!\psi(p_2)
\end{eqnarray}
where
\begin{eqnarray}
I_1&=&\int_q{1\over 2\mu+\bar V\cdot (l_3+q)}
{1\over 2\mu+V\cdot (l_4-q)}{1\over(q+l)^2q^2}
\label{integral1} \\
I_2&=&\int_q{1\over 2\mu+\bar V\cdot (l_1-q)}
{1\over 2\mu+V\cdot (l_4-q)}{1\over(q+l)^2q^2}
\label{integral2}
\end{eqnarray}
In the limit that the residual momenta $l_i\to0$, we find
$I_1=I_2$, if we rotate the $q_0$ axis into $\vec v_F\cdot\vec q$
axis, and thus the color and Lorentz structures of the amplitudes
in both theories are same.
To perform the integration, we take, for convenience,
the external residual
momenta are perpendicular to the Fermi velocity, $V\cdot l_i=0$.
Then, we get for $l_i/\mu\to0$
\begin{eqnarray}
I_1&=&\int_0^1{\rm d}x\int_0^{1-x}{\rm d}y\int_q
{-2\over\left[
(1-x-y)q^2+x(q+l)^2+yq_0^2-y(\vec v_F\cdot\vec q+2\mu)^2
\right]^3}\\
&=&{i\over 64\pi^2\mu^2}\left[2-\ln\left({-l^2\over4\mu^2}\right)
+O(l^2/\mu^2)\right].
\nonumber
\end{eqnarray}
Note that the scattering amplitude $A_4^{\rm QCD}$ is IR
divergent as $l_i\to0$, which is the same infrared divergence of
the amplitude in the effective theory, Eq.~(\ref{amp}), as it should
be because the low energy effective theory is equivalent to
the microscopic theory in the infrared limit by construction.
Since both amplitudes have to
be same at the matching scale $\Lambda=\mu$, we need a four-quark
operator in the effective theory, given by the difference of the
amplitudes at the matching scale $\mu$:
\begin{eqnarray}
A_4^{\rm QCD}-A_4^{\rm eff}
&=&{ig_s^4\over 64\pi^2\mu^2}\left({1\over\epsilon}-\gamma-4\right)
(2\pi)^4\delta^4(\sum_i{}^{\prime}p_i)
\psi^{\dagger}(p_3)\gamma^{\mu}_{\perp}\gamma^{\nu}_{\perp}T^aT^b
\psi(p_1)    \nonumber\\
& & \times \psi^{\dagger}(p_4)\left(\gamma_{\mu}\gamma_{\nu}
T^aT^b+\gamma_{\nu}\gamma_{\mu}T^bT^a \right)
\psi(p_2)+O(l_i/\mu) \\
&=& \left<p_3,p_4\right|i\int {\rm d}^4x~{\cal L}_{4f}^1
\left|p_1,p_2\right>+O(l_i/\mu), \nonumber
\end{eqnarray}
where, at the matching scale $\mu$,
the new effective four-quark operator~\footnote{
If we had matched a quark-quark scattering amplitude with
momenta not opposite to each other, we would get similarly a
four-quark operator in the effective Lagrangian. But, since it is
irrelevant as we scale toward the Fermi sea, it will not affect
the low energy dynamics significantly.}
generated at one-loop is given as
\begin{eqnarray}
{\cal L}_{4f}^1=-{g^{\rm bare}_1\over 2\mu^2}
\sum_{\vec v_F}& &\!\!\!\!\!\!\left[\psi^{\dagger}(\vec v_F,x)
\gamma^{\mu}_{\perp}\gamma^{\nu}_{\perp}T^aT^b
\psi(\vec v_F,x)\right.\label{amp1}\\
&\times&\left.
\psi^{\dagger}(-\vec v_F,x)
\left(\gamma_{\mu}\gamma_{\nu}
T^aT^b+\gamma_{\nu}\gamma_{\mu}T^bT^a \right)
\psi(-\vec v_F,x) \right].
\nonumber
\end{eqnarray}
In the modified minimal subtraction, the renormalized
four-quark coupling is $g_1^{\rm ren}=2\alpha_s^2$
at the matching scale $\mu$.
It is convenient to rewrite the marginal four-quark operator
by Fierz-transforming
the product of gamma matrices and $SU(3)_C$
generators in the amplitude Eq.~(\ref{amp1}).
Using $\left(T^a\right)_{tu}\left(T^a\right)_{vs}=
1/2\delta_{ts}\delta_{uv}-1/6\delta_{tu}\delta_{vs}$ and the Fierz
transformation,
we get
\begin{eqnarray}
& & \left[P_+
\gamma^{\mu}_{\perp}\gamma^{\nu}_{\perp}P_+\right]_{ji}
\left(T^aT^b\right)_{ut}\cdot
\left[P_-\left(\gamma_{\mu\perp}\gamma_{\nu\perp}T^aT^b
+\gamma_{\nu\perp}\gamma_{\mu\perp}T^bT^a\right)P_-\right]_{lm;vs}
\nonumber\\
& & =\left[P_+\gamma^{\mu}_{\perp}\gamma^{\nu}_{\perp}P_+\right]_{ji}
\left(T^aT^b\right)_{ut}\cdot \left[ g_{\mu\nu}P_-
\left({1\over3}\delta^{ab}+d^{abc}T^c
\right)+\left(P_-\sigma_{\mu\nu}P_-\right)f^{abc}
T^c\right]_{lm,vs} \nonumber \\
& & =
\left(P_+\right)_{ji}\left(P_-\right)_{lm}\left(
{11\over18}\delta_{ut}\delta_{vs}+{5\over6}\delta_{us}\delta_{tv}
\right) +\left(P_+\gamma_5\right)_{ji}
\left(P_-\gamma_5\right)_{lm}\left(
-{1\over2}\delta_{ut}\delta_{vs}+{3\over2}\delta_{us}\delta_{tv}
\right)
\\
& &=-\left(P_+\right)_{ji}\left(P_-\right)_{lm}
\left({\sqrt{2}\over9}\delta^A_{uv;ts}
-{13\over9\sqrt{2}}\delta^S_{uv;ts}\right)
-\left(P_+\gamma_5\right)_{ji}\left(P_-\gamma_5\right)_{lm}
\left(
\sqrt{2}\delta^A_{uv;ts}-{1\over\sqrt{2}}\delta^S_{uv;ts}
\right),
\nonumber
\end{eqnarray}
where $i,j,l,m$ denote the Dirac indices and $t,s,u,v$ the color
indices.
In the last line, the color indices in the operator
are further arranged to decompose the amplitude into
the irreducible representations of $SU(3)_C$ by introducing
invariant tensors in color space,
$\delta^S_{uv;ts}\equiv\left(
\delta_{ut}\delta_{vs}+\delta_{us}\delta_{vt}\right)
/\sqrt{2}$
and
$\delta^A_{uv;ts}\equiv\left(
\delta_{ut}\delta_{vs}-\delta_{us}\delta_{vt}\right)/
\sqrt{2}$.

The four-quark operator in the effective
theory becomes then
\begin{eqnarray}
{\cal L}_{4f}^1&=&{1\over 2\mu^2}\sum_{\vec v_F}
\left[g_{us;tv}\psi^{\dagger}_t(\vec v_F,x)\psi_s(\vec v_F,x)
\psi^{\dagger}_v(-\vec v_F,x)\psi_u(-\vec v_F,x) \right. \\
& & \left.\quad\quad
+h_{us;tv}\psi^{\dagger}_t(\vec v_F,x)\gamma_5\psi_s(\vec v_F,x)
\psi^{\dagger}_v(-\vec v_F,x)\gamma_5\psi_u(-\vec v_F,x)
\right]
\nonumber
\end{eqnarray}
with
$g_{us;tv}=g_{\bar3}\delta^A_{us;tv}-g_6\delta^S_{us;tv}$,
$h_{us;tv}=h_{\bar3}\delta^A_{us;tv}-h_6\delta^S_{us.s;tv}$.
The value of couplings at the matching scale $\mu$ are given as
$g_{\bar3}=4\sqrt{2}\alpha_s^2/9=2g_6/(13)$
and $h_{\bar3}=4\sqrt{2}\alpha_s^2=2h_6$.

\section{screening mass}
\label{sec:label4}

As discussed in~\cite{kapusta88,manuel,nair,kapusta},
the quark loop correction
to the vacuum polarization tensor gives rise to color screening in
quark-gluon plasma, while the gluon loop renormalizes the color
charge.
We first calculate the vacuum polarization tensor in QCD and then
compare it with the one in the effective theory when we perform the
one-loop matching for the gluon two-point amplitude.

Since the Feynman propagator for quarks of mass $m$ in matter
at zero temperature is given by
\begin{equation}
iS_F(x)=\int {{\rm d}^4p\over (2\pi)^4} e^{-ip\cdot x}{i\over
(1+i\epsilon)p_0\gamma^0-\vec p\cdot \vec \gamma+\mu\gamma^0-m},
\label{fullp}
\end{equation}
we find, performing the integration over $p_0$,
\begin{eqnarray}
iS_F(x)&=&
\theta(x_0)\int {{\rm d}^3p\over (2\pi)^3}{\not\! p+m\over
2p_0} \theta(p_0-\mu)e^{-ip\cdot x+i\mu x_0}\nonumber\\
&-&\theta(-x_0)\int {{\rm d}^3p\over (2\pi)^3}
\left[{\not\!p+m\over 2p_0}\theta(\mu-p_0)e^{-ip\cdot x+i\mu x_0}
+{\not\!p-m\over 2p_0}e^{ip\cdot x+i\mu x_0}
\right],
\end{eqnarray}
where $p_0=\sqrt{\left|\vec p\right|^2+m^2}$.  We see that this
propagator agrees with the one obtained by the canonical
quantization~\cite{manuel} except the overall phase factor due to the
shift in the energy by $\mu$ to set the energy of the Fermi surface
to be zero.

In terms of this full propagator Eq.~(\ref{fullp}), the
quark-loop contribution to the one-loop vacuum polarization tensor
becomes
\begin{eqnarray}
\Pi^{\mu\nu}_{ab\rm full}(p)&=& g_s^2\int {\rm d}^4x e^{-ip\cdot
x}\left< J_a^{\mu}(x)J_b^{\nu}(0)\right>\\
&=&{iN_f\over2}g_s^2\delta_{ab}\int{\rm d^3q\over (2\pi)^3}{1\over
2q_0}{1\over 2k_0} \left[ {\theta(q_0-\mu)\theta(\mu-k_0)\over
q_0-k_0+p_0-i\epsilon} T^{\mu\nu}(q,k)
-{\theta(\mu-q_0)\theta(k_0-\mu)\over
q_0-k_0+p_0+i\epsilon}T^{\mu\nu}(q,k)\right.\nonumber\\
&&\left.
+{\theta(q_0-\mu)\over
q_0+k_0+p_0-i\epsilon}T^{\mu\nu}(q,k^{\prime})
-{\theta(k_0-\mu)\over-q_0-k_0+p_0+i\epsilon}T^{\mu\nu}(q,k^{\prime
\prime})\right],
\label{vacpol}
\end{eqnarray}
where $T^{\mu\nu}(q,k)={\rm
Tr}\left[\gamma^{\mu}\!\not\!q\gamma^{\nu}\!\not\!k\right]$,
$\vec k=\vec q+\vec p$, $\vec k^{\prime}=-\vec q-\vec p$,
and $\vec k^{\prime\prime}=-\vec q+\vec p$.

If we rewrite the step functions as
$\theta(q_0-\mu)=1-\theta(\mu-q_0)$
and $\theta(k_0-\mu)=1-\theta(\mu-k_0)$ for the last two terms in
Eq.~(\ref{vacpol}), one can easily see
that the quark-loop contribution of the vacuum polarization tensor
consists of two parts, one due to the matter and the other due to the
vacuum,
$\Pi^{\mu\nu}_{ab\rm full}(p) =\Pi^{\mu\nu}_{ab\rm
mat}(p)+\Pi^{\mu\nu}_{ab\rm vac}(p) $, where $\Pi^{\mu\nu}_{ab\rm
vac}$ is the quark-loop contribution when there is no matter ({\it
i.e.} $\mu=0$),
\begin{equation}
\Pi^{\mu\nu}_{ab\rm vac}={iN_f\over2}g_s^2\delta_{ab}
\int{\rm d^3q\over (2\pi)^3}{1\over 2q_0}{1\over 2k_0}
\left[{T^{\mu\nu}(q,k^{\prime})\over q_0+k_0+p_0-i\epsilon}
-{T^{\mu\nu}(q,k^{\prime\prime})\over-q_0-k_0+p_0+i\epsilon}
\right],
\end{equation}
where $N_f$ is the number of light quark flavors.
For $\left|p^{\mu}\right|\ll\mu$,
the matter part of the vacuum
polarization becomes,  with $M^2=N_f
g_s^2\mu^2/(2\pi^2)$,
\begin{equation}
\Pi^{\mu\nu}_{ab\rm mat}
=-{iM^2\over2}\delta_{ab} \int{{\rm d}\Omega_{\vec v_F}\over 4\pi}
\left({-2\vec p\cdot\vec v_FV^{\mu}V^{\nu}\over p\cdot V
+i\epsilon \vec p\cdot\vec v_F}+g^{\mu\nu}-{
V^{\mu}{\bar V}^{\nu}+{\bar V}^{\mu}V^{\nu}\over2}\right)
\end{equation}
which is transversal,
$p_{\mu}\Pi^{\mu\nu}_{ab\rm mat}(p)=0$~\cite{manuel}.

Now, we calculate the quark contribution to the one-loop vacuum
polarization tensor in the effective theory.
Quarks in the (high density) effective theory are
almost on-shell at low energy and the quark current, which consists
of states near the Fermi surface, is proportional to its velocity,
$V^{\mu}=(1,\vec v_F)$:
\begin{equation}
J^{\mu a}(x)\equiv\sum_{\vec v_F}\bar\psi_+(\vec
v_F,x)\gamma^{\mu}T^a\psi_+(\vec v_F,x)=\sum_{\vec v_F}V^{\mu}
\bar\psi_+(\vec v_F,x)\gamma^0T^a\psi_+(\vec v_F,x),
\label{current}
\end{equation}
where we used $P_+\psi_+(\vec v_F,x)=\psi_+(\vec v_F,x)$ and
$P_-\gamma^{\mu}P_+=P_-\gamma^0 V^{\mu}$.
Therefore, the gluons transversal to the quark velocity is decoupled
at low energy and the gluons longitudinal to the quark velocity is
coupled to the quark color charge density in a combination given as
\begin{equation}
A^a_{\mu}(x)J^{\mu a}(x)=\sum_{\vec v_F}\bar\psi_+(\vec v_F,x)\gamma^0
T^a\psi_+(\vec v_F,x)~V\cdot A^a(x).
\label{min}
\end{equation}
Therefore, in the leading order in $1/\mu$ expansion, the quark-gluon
coupling does not involve the spin of quarks like a scalar coupling.
For a given quark, moving nearly at a Fermi momentum $\mu\vec v_F$,
only one component of gluons, namely $A_0-\vec v_F\cdot \vec A$
combination, couples to the quark in the leading order.

With the quark current given in Eq.~(\ref{current}), we obtain the
quark-loop contribution to the vacuum polarization in the effective
theory as
\begin{eqnarray}
\Pi^{\mu\nu}_{ab}(p)&=&g_s^2\int {\rm d}^4x~e^{-ip\cdot x}
\sum_{\vec v_F}
\left<0|TJ^{\mu}_a(\vec v_F,x)J^{\nu}_b(\vec
v_F^,0)\right>\\
&=&-{g_s^2}N_f\delta_{ab}
\sum_{\vec v_F}V^{\mu}V^{\nu}
\int {{\rm d}^4q\over (2\pi)^4}
{\rm Tr}
\left[\left({1-\vec \alpha\cdot\vec v_F\over2}\right)
\gamma^0{i\over \not q_{\parallel}+i\epsilon}
\gamma^0{i\over \not q_{\parallel}+
\not p_{\parallel}+i\epsilon}\right] \nonumber \\
&=&-{i\over8\pi}\delta_{ab}\sum_{\vec v_F}V^{\mu}V^{\nu} \left[
1-{(p^0+\vec p\cdot\vec v_F)^2\over p_{\parallel}^2+i\epsilon}
\right]M^2,
\label{vpol}
\end{eqnarray}
where
$p_{\parallel}^{\mu}=(p^0,\vec v_F \vec
v_F\cdot \vec p)$.
We note that the quark-loop contribution to the vacuum polarization
tensor in the effective theory is exactly same as the sum of the
first and second terms in
the QCD vacuum polarization, Eq.~(\ref{vacpol}). This is what it
should be, since the effective theory does reproduce the
contribution by quarks and holes in QCD when the external gluon
momentum $p\to0$. The third and fourth terms in
Eq.~(\ref{vacpol}), due to the quark and anti-quark pair creation,
is absent in the effective theory because the anti-quarks are
integrated out. The effect of anti-quarks will be taken into account
when we match the gluon two-point amplitudes.

To match the gluon two-point amplitudes in both theories,
we therefore need to add a term in the one-loop
effective Lagrangian,
\begin{equation}
{\cal L}_{\rm eff}\ni -{M^2\over 16\pi}\sum_{\vec v_F}A_{\perp}^{a\mu}
A_{\perp\mu}^a,
\label{add}
\end{equation}
which also ensures the gauge invariance of the effective Lagrangian
at one-loop.
The static screening mass can be read off from the vacuum polarization
tensor in the limit $p_0\to0$, which is in the effective theory
\begin{equation}
\Pi^{\mu\nu}_{ab}(p_0\to0,\vec p)\simeq
-iM^2\delta_{ab}\delta^{\mu0}\delta^{\nu0}.
\end{equation}
We see that the electric gluons, $A_0^a$, have a screening mass, $M$, but
the static magnetic gluons are not screened at one-loop  due to the
added term Eq.~(\ref{add}),  which holds at all orders in perturbation
as in the finite temperature~\cite{kapusta,pisarski}.

Finally, by matching the quark two-point amplitudes,
we get a one-loop low-energy (Wilsonian) effective Lagrangian
density;
\begin{eqnarray}
{\cal L}_{\rm eff}&=&-{1\over4}(1+a_1)
\left(F_{\mu\nu}^a\right)^2 -
{M^2\over 16\pi}A_{\perp}^{a\mu}A_{\perp\mu}^a
+(1+b_1)\bar\psi i\gamma_{\parallel}^{\mu}D_{\mu}\psi
-{1\over2\mu}(1+c_1)
\psi^{\dagger}\left(\gamma_{\perp}\cdot D\right)^2 \psi
\nonumber \\
&+&
{1\over2\mu^2}
\left[
\left(g_{\bar 3}\delta^A_{us;tv}-g_6\delta^S_{us;tv}\right)
\psi^{\dagger}_t(\vec v_F,x)\psi_s(\vec v_F,x)
\psi^{\dagger}_v(-\vec v_F,x)\psi_u(-\vec v_F,x)
\right. \label{effr}
\\
& &\left.
+\left(h_{\bar 3}\delta^A_{us;tv}-h_6\delta^S_{us;tv}\right)
\psi^{\dagger}_t(\vec v_F,x)\gamma_5\psi_s(\vec v_F,x)
\psi^{\dagger}_v(-\vec v_F,x)
\gamma_5\psi_u(-\vec v_F,x)\right]+\cdots,
\nonumber
\end{eqnarray}
where the summation over $\vec v_F$ is suppressed
and the coefficients
$a_1,b_1,c_1$ are dimensionless and of order
$\alpha_s(\mu)$. The ellipsis denotes the irrelevant
four-quark operators and terms with more external fields and
derivatives.

\section{RG analysis and  Gap equation}
\label{sec:label6}

As we scale further down, the effective four-quark
operators will evolve together with other operators, which can be seen
by further integrating out the high frequency modes,
$s\mu<|l_{\mu}|<\mu$. The scale dependence of the four-quark operators
has three pieces.
One is from the one-loop matching condition for the
four-quark amplitudes and the other two are from the loop corrections
to the four-quark operators.
Putting all contribution together, we find the one-loop
renormalization group equations for the four-quark
operators~\cite{hong98} to be
\begin{equation}
s{\partial \over\partial s}{\bar g}_i=-
\gamma_i\alpha_s^2-{1\over 4\pi^2}{\bar g}_i^2-
{\ln2\over12\pi}\delta_i{\bar g}_i\alpha_s,
\label{rge}
\end{equation}
where $i=(6,{\bar 3})$, $\bar g_6=-g_6$,
$\bar g_{\bar3}=g_{\bar3}$ and
$\gamma_i=(\sqrt{2}/9)(13/4,1/2)$
and $\delta_i=(-1,2)$.
Since in the high density limit the quark-gluon coupling is
(1+1)-dimensional, the quarks do not contribute to
the running of the strong coupling. The one-loop
$\beta$ function for
the strong coupling constant at high density is
$\beta(\alpha_s)=-11/(2\pi)\alpha_s^2$.

To solve the RG equation, Eq.~(\ref{rge}),
we introduce a new variable
\begin{equation}
y_i\equiv {1\over 22\pi\alpha_s(\mu)}
\left(\bar g_i+{\pi\ln2\over6}\delta_i\alpha_s(\Lambda)\right)
\quad{\rm and}\quad t={11\over2\pi}\alpha_s(\mu)\ln s
\end{equation}
Then, the RG equations becomes
\begin{equation}
{{\rm d}y_i\over {\rm d}t}=-y_i^2-{a_i\over (1+t)^2},
\end{equation}
where $a_i=(132\delta_i\ln2+144\gamma_i-\delta_i^2)/(17424)$.
Further, letting $y_i=f_i(t)/(1+t)$, we get
\begin{equation}
(1+t){{\rm d}f_i\over {\rm d}t}=-\left(f_i^2-f_i+a_i\right).
\label{rge2}
\end{equation}
Integrating Eq.~(\ref{rge2}), we get
\begin{equation}
f_i(t)={C_2(f_i(0)-C_1)(1+t)^{-(C_1-C_2)}-C_1(f_i(0)-C_2)
\over (f_i(0)-C_1)(1+t)^{-(C_1-C_2)}-(f_i(0)-C_2)},
\end{equation}
where $C_1=(1+\sqrt{1-4a_i})/2=1-C_2$ and $f_i(0)=\left(g_i(\mu)
+\pi\ln2\delta_i\alpha_s(\mu)/6\right)/(22\pi\alpha_s(\mu))$.
Since $a_i\ll1$, we take $C_1\simeq1$, $C_2\ll1$. And also,
for $\mu\gg\Lambda_{\rm QCD}$, $f_i(0)\simeq
\delta_i\ln2/(132)\ll1$. Therefore, for $1+t\to0^+$ or
$s\to\exp\left[-2\pi/(11\alpha_s(\mu))\right]$,
we get $\bar g_i(t)+\pi\ln2\delta_i\alpha_s(t)/6
\simeq22\pi C_2\alpha_s(t)$.
Namely,
\begin{eqnarray}
{\bar g}_{i}(\Lambda)\simeq{2\pi\over
11}\alpha_s(\Lambda)\left[\gamma_i
- {(\ln2)^2\over144}\delta_i^2\right].
\end{eqnarray}
At a scale much less than the chemical potential,
$\Lambda\ll\mu$, $g_6(\Lambda)\simeq 0.29\alpha_s(\Lambda)$ and
$g_{\bar3}(\Lambda)\simeq 0.04\alpha_s(\Lambda)$.
Similarly for $h_i$, $\gamma_i=(\sqrt{2}/2)(2,1)$ and
$\delta_i=(-1,2)$ and we get
$h_6(\Lambda)=0.81\alpha_s(\Lambda)$ and
$h_{\bar3}(\Lambda)\simeq 0.4\alpha_s(\Lambda)$.


At a scale below the screening mass, we further integrate out
the electric gluons, which will
generate four-quark interactions.
For quarks moving with opposite Fermi momenta, the electric-gluon
exchange four-quark interaction is given as
\begin{equation}
{\cal L}_{1g}\ni -{g_s^2(M)\over 2M^2}\sum_{\vec v_F}
\bar\psi\gamma^{0}T_a\psi(\vec v_F,x)
\bar\psi\gamma_0T_a\psi(-\vec v_F,x).
\label{gluon}
\end{equation}
Using  $T^a_{tu}T^a_{vs}
=1/2\delta_{ts}\delta_{uv}-1/6\delta_{tu}\delta_{vs}$,
we find that the four-quark couplings are shifted as
$g_{\bar3}(M)\to g_{\bar3}(M)+2\sqrt{2}g_s^2(M)/3\simeq
0.95g_s^2(M)$
and $g_6(M)\to g_6(M)+\sqrt{2}g_s^2(M)/3\simeq0.49g_s^2(M)$,
while $h_i$'s are unchanged since Eq.~(\ref{gluon}) does not
involve $\gamma_5$.

As we approach further to the Fermi surface, closer than the
screening mass $M$, the four-quark operators in the color anti-triplet
channel become stronger, because the $\beta$ function for the
attractive  four-quark operators is negative,
$\beta (g_{\bar 3})=-g_{\bar3}^2/(4\pi^2)$.
If the four-quark interaction is dominant at low energy, it leads to
vacuum instability in the infrared region  by forming a color
anti-triplet condensate or Cooper pair,
But, since the long-range color-magnetic interactions also
become  strong at low energy, we need to consider both
interactions to determine the Cooper-pair gap.

Since both relevant interactions are attractive for a pair of quarks
in color anti-triplet channel with opposite Fermi momenta, they
may
lead to condensates of quark pairs in color anti-triplet channel.
To describe the Cooper-pair gap equation, we introduce a charge
conjugate field,
\begin{equation}
\left(\psi_c\right)_{i}(\vec v_F,x)=C_{ij}\bar\psi_{j}(-\vec v_F,x),
\end{equation}
where $i$ and $j$ are Dirac indices and the matrix $C$
satisfies $C^{-1}\gamma_{\mu}C=-\gamma_{\mu}^T$.
Then, we can write the inverse propagator for $\Psi(\vec
v_F,x)\equiv(\psi(\vec v_F,x),\psi_c(\vec v_F,x))^T$ as
\begin{equation}
S^{-1}(\vec v_F,l)= \gamma_0\pmatrix{Z(l_{\parallel})l\cdot V &
       -\Delta(l_{\parallel}) \cr
-\Delta^{\dagger}(l_{\parallel})& Z(l_{\parallel})l\cdot\bar V\cr},
\end{equation}
where $Z(l_{\parallel})$ is the wave function renormalization constant
and $\Delta(l_{\parallel})$ is the Cooper-pair gap. The
Cooper-pair gap is nothing but a Majorana mass for quarks. When
Cooper-pairs form and Bose-condense, the quarks get a Majorana mass
dynamically and the Fermi sea opens up a gap.

Schwinger-Dyson (SD) equations are infinitely coupled integral
equations for Green functions. In order to solve the Schwinger-Dyson
equations, we
need to truncate them consistently.
Since $g_s(M)$ is still weak
at high density $M\gg\Lambda_{\rm QCD}$, one may use the ladder
approximation which is basically the weak coupling expansion of
the SD equations, consistently with the Ward-Takahashi identity. Another
gauge-invarinat truncation is so-called hard-dense-loop (HDL)
re-summation, which can be derived gauge invariantly from the transport
equation~\cite{manuel} as in the finite temperature~\cite{braaten}.
When the gluon momentum is of order of the
screening mass $M\sim g_s\mu$, the quark one-loop vacuum
polarization is also same order as we can see from
Eq.~(\ref{vpol}). Therefore, if the important momentum scale of
the diagrams is of order of $M$, one needs to re-sum all the bubble
diagrams to get the consistent gluon propagator.
Since the HDL re-summed gluon propagator correctly incorporates
the medium effects like screening at high density, we will solve the
SD equations for the quark propagator in the HDL approximation
to calculate the Cooper-pair gap.

In the HDL
approximation, the SD  equations for
the quark propagator are given as, neglecting small $h_i$
four-Fermi couplings,
\begin{eqnarray}
\left[Z(p_{\parallel})-1\right]p\cdot V
&=& (-ig_s)^2
\int_lV^{\mu}D_{\mu\nu}(p-l){\bar V}^{\nu}
{T^aZ(l_{\parallel})l\cdot V {T^a}^T\over Z^2
l_{\parallel}^2-\Delta(l_{\parallel})^2} +{g_{\bar3}\over
\mu^2}\int_l  {Z(l_{\parallel})
l\cdot V\over
Z^2l_{\parallel}^2-\Delta(l_{\parallel})^2}
\\
\Delta(p_{\parallel})&=&(-ig_s)^2
\int_lV^{\mu}D_{\mu\nu}(p-l){\bar V}^{\nu}
{T^a\Delta(l_{\parallel}){T^a}^T\over  Z^2
l_{\parallel}^2-\Delta(l_{\parallel})^2} +{g_{\bar3}\over
\mu^2}\int_l  {\Delta(l_{\parallel})\over
Z^2l_{\parallel}^2-\Delta(l_{\parallel})^2},
\end{eqnarray}
where the gluon propagator $D_{\mu\nu}$ is given
in the HDL approximation
as,\footnote{In the Schwinger-Dyson equation the loop
momentum should take the whole range up to the ultraviolet cutoff,
which is the chemical potential $\mu$ in the case of high density effective
theory. Hence the gluon propagator includes both magnetic and electric
gluons.} following the notations used by Sch\"afer and Wilczek~\cite{sw99},
\begin{equation}
iD_{\mu\nu}(k)={P_{\mu\nu}^{T}\over k^2-G}+{P^L_{\mu\nu}\over k^2-F}
-\xi {k_{\mu}k_{\nu}\over k^4},
\end{equation}
where $\xi$ is the gauge parameter and
the projectors are defined by
\begin{eqnarray}
P^T_{ij}&=&\delta_{ij}-{k_ik_j\over |\vec k|^2},
\quad P_{00}^T=0=P_{0i}^T\\
P^L_{\mu\nu}&=&-g_{\mu\nu}+{k_{\mu}k_{\nu}\over k^2}-P^T_{\mu\nu}.
\end{eqnarray}

Before solving the SD equations, we first show that
the preferred solution of the SD equations supports the color-flavor
locking condensate predicted in~\cite{wilczek2}.
Since the gluon interaction is vectorial, the gluon exchange
interaction in the gap equation does not distinguish the handedness
of quarks and thus it will generate same condensates regardless of
handedness;
$\left|\left<\psi_L\psi_L\right>\right|=
\left|\left<\psi_R\psi_R\right>\right|
=\left|\left<\psi_L\psi_R\right>\right|$, suppressing other
quantum numbers. But, the four-Fermi
interaction in the effective Lagrangian, Eq.~(\ref{effr}), is chiral,
which can be seen if we rewrite it
in the handedness basis as following;
\begin{eqnarray}
{\cal L}_{\rm 4f}^1&\ni&
{g_{\bar3}+h_{\bar3}\over 2\mu^2}\delta^A_{us;tv}\left[
{\psi^{\dagger}_L}_t(\vec v_F,x){\psi_L}_s(\vec v_F,x)
{\psi^{\dagger}_L}_v(-\vec v_F,x){\psi_L}_u(-\vec v_F,x)
+\left(L\leftrightarrow R\right)
\right]\\
&+&
{g_{\bar3}-h_{\bar3}\over 2\mu^2}\delta^A_{us;tv}\left[
{\psi^{\dagger}_L}_t(\vec v_F,x){\psi_L}_s(\vec v_F,x)
{\psi^{\dagger}_R}_v(-\vec v_F,x){\psi_R}_u(-\vec v_F,x)
+\left(L\leftrightarrow R\right)
\right].
\nonumber
\end{eqnarray}
We see that the $LL$ (or $RR$) four-Fermi coupling is bigger than the
$LR$ four-Fermi coupling, since $h_{\bar3}>0$.
Therefore, the gap in $LL$ or $RR$ channel
will be bigger than the one in $LR$ channel due to the difference in
the four-Fermi couplings. Thus, the $LL$ or $RR$
condensate is energetically more preferred than the $LR$ condensate.
We also note that in the effective theory the gluons are blind not
only to flavors but also to the Dirac indices of quarks
because they couple to quark currents in the combination of
$V\cdot A$ like a scalar field, as given in Eq.~(\ref{min}).
Therefore, in the SD equations,
the diquark Cooper-pair can be decomposed into color anti-triplet
and color sextet. But, since quarks of opposite momenta are
attractive in the anti-triplet channel while repulsive in the
sextet channel, without loss of generality, we can
write the Cooper-pair gap in color anti-triplet~\footnote{
At high but finite density, the Cooper-pair gap contains a
small component of color-sextet~\cite{PR99}.
But we will ignore this for simplicity.}.
Since quarks are anti-commuting, the
only possible way to form  diquark (S-wave) condensate is either in
spin-singlet or in spin-triplet:
\begin{eqnarray}
\left<{\psi_L}^a_{i\alpha}(\vec v_F,x){\psi_L}^b_{j\beta}(-\vec v_F,x)
\right>&=&-\left<{\psi_R}^a_{i\alpha}(\vec v_F,x)
{\psi_R}^b_{j\beta}(-\vec v_F,x)\right>\\
&=&\epsilon_{ij}\epsilon^{abc}
K_{\left[\alpha\beta\right]c}(p_F)+
\delta_{ij}\epsilon^{abc}
K_{\left\{\alpha\beta\right\}c}(p_F),
\end{eqnarray}
where $a,b,c=1,2,3$ are color indices,
$\alpha,\beta,\gamma=u,d,s,\cdots,N_f$ flavor indices, and
$i,j=1,2$ spinor indices. Indices in the bracket and in the
curled bracket are anti-symmetrized and symmetrized, respectively.
But, the spin-one component of the gap, $K_{\{\alpha\beta\}c}$,
vanishes algebraically, since $\psi(\vec v_F,x)=1/2\left(1+
\vec\alpha\cdot\vec v_F\right)\psi(\vec v_F,x)$ and
$(1+\vec\alpha\cdot\vec v_F)_{il}(1-\vec\alpha\cdot\vec v_F)_{lj}=0$.

When $N_f=3$, the spin-zero component of
the condensate  becomes (flavor)
anti-triplet,
\begin{equation}
K_{\left[\alpha\beta\right]c}(p_F)
=\epsilon_{\alpha\beta\gamma}K_c^{\gamma}(p_F).
\end{equation}
Using the global color and flavor symmetry, one can
always diagonalize the spin-zero condensate as
$K_c^{\gamma}=\delta_c^{\gamma}K_{\gamma}$.
To determine the parameters, $K_u$, $K_d$, and $K_s$,
we need to minimize the vacuum energy for the
condensate. By the Cornwall-Jackiw-Tomboulis
formalism~\cite{jackiw74},
the vacuum energy in the HDL approximation is given as
\begin{eqnarray}
V(\Delta)&=& -{\rm Tr}~\ln S^{-1}+{\rm Tr}~\ln\not\!\partial
+{\rm Tr}~(S^{-1}-\not\!\partial)S+({\rm 2PI~diagrams})
\nonumber\\
&=&{\mu^2\over4\pi}\sum_{i=1}^9\int {d^2l_{\parallel}\over
(2\pi)^2}\left[\ln\left({l_{\parallel}^2\over
l_{\parallel}^2+\Delta_{i}^2(l_{\parallel})} \right)+
{1\over2}\cdot{\Delta_{i}^2(l_{\parallel})\over
l_{\parallel}^2+\Delta_{i}^2(l_{\parallel})}\right]+h.o.,
\label{ve}
\end{eqnarray}
where $h.o.$ are the higher order terms in the HDL approximation,
containing more powers of coupling $g_s$, and
$\Delta_i$'s are the eigenvalues of color anti-symmetric
and flavor anti-symmetric $9\times9$ gap, $\Delta_{\alpha\beta}^{ab}$.
The 2PI diagrams are two-particle-irreducible vacuum diagrams. There
is only one such diagram (see Fig.~3) in the leading order
HDL approximation.

Since the gap depends only on energy in the leading order,
one can easily perform the momentum integration in
(\ref{ve}) to get\footnote{
If the condensate forms, the vacuum energy due to the gluons also
depends on the gap due to the Meisner effect. But, it turns
out to be subleading, compared to the quark vacuum energy;
$V_g(\Delta)\sim M^2\Delta^2
\ln(\Delta/\mu)\sim g_s\mu^2\Delta^2$\cite{ehhs99}.},
\begin{eqnarray}
V(\Delta)&=&{\mu^2\over 4\pi^2}\int_0^{\infty}
dl_0\left(-{\Delta_i^2\over
\sqrt{l_0^2}+\sqrt{l_0^2+\Delta_i^2}}+{1\over4}\cdot
{\Delta_i^2\over\sqrt{l_0^2+\Delta_i^2}}
\right)\nonumber\\
&\simeq&
-0.43 {\mu^2\over 4\pi^2}
    \sum_i\left|\Delta_i(0)\right|^2,
\end{eqnarray}
where in the second line we used an approximation that
\begin{equation}
\Delta_i(l_0)\simeq
\left\{
\begin{array}{ll}
\Delta_i(0) & \mbox{if $\left|l_0\right|<\left|\Delta_i(0)\right|$},\\
0 & \mbox{otherwise}.
\end{array}
\right.
\end{equation}
Were $\Delta_i$ independent of each other, the global minimum
should occur at $\Delta_i(0)={\rm const.}$ for all $i=1,\cdots,9$.
But, due to the global
color and flavor symmetry, only three of them are independent.
Similarly to the condensate,
the gap can be also diagonalized by the color and flavor symmetry as
\begin{equation}
\Delta^{\alpha\beta}_{ab}=\epsilon_{\alpha\beta\gamma}
\epsilon^{abc}\Delta_{\gamma}\delta^{\gamma}_c.
\end{equation}
Without loss of generality, we can take $\left|\Delta_u\right|
\ge \left|\Delta_d\right|\ge\left|\Delta_s\right|$.
Let $\Delta_d/\Delta_u=x$ and $\Delta_s/\Delta_u=y$.
Then, the vacuum energy becomes
\begin{equation}
V(\Delta)\simeq -0.43 {\mu^2\over 4\pi^2}\left|\Delta_u\right|^2f(x,y),
\end{equation}
where $f(x,y)$ is a complicate function of $-1\le x,y\le 1$
that has a maximum at $x=1=y$, $f(x,y)\le 13.4$.
Therefore, the vacuum energy has a global minimum when
$\Delta_u=\Delta_d=\Delta_s$, or in terms of the eigenvalues of the gap
\begin{equation}
\Delta_i=\Delta_u \cdot(1,1,1,-1,1,-1,1,-1,-2).
\end{equation}
Among nine quarks, $\psi_a^{\alpha}$, eight have (Majorana) mass
$\Delta_u$ and one has mass $2\Delta_u$.

Since the condensate is
related to the off-diagonal component of the quark propagator at high
momentum as, suppressing the color and flavor indices,
\begin{eqnarray}
\left<\psi(\vec v_F,x)\psi(-\vec v_F,x)\right>
&\sim&\lim_{y\to x}\int {{\rm d}^4l\over (2\pi)^4}e^{il\cdot (x-y)}
{\Delta(l_{\parallel})\over l_{\parallel}^2-\Delta^2(l_{\parallel})}
\nonumber\\
&=&\lim_{y\to x}\left[\delta^2(\vec x_{\perp}-\vec y_{\perp})
{\Delta(0)\over4\pi^2|x_{\parallel}-y_{\parallel}|^{\gamma_m}}
+\cdots\right],
\end{eqnarray}
where $\gamma_m$ is the anomalous dimension of the condensate and
the ellipsis are less singular terms.
Being proportional to the gap, the condensate is
diagonalized in the basis where the gap is diagonalized.
Thus, we have shown that in the HDL approximation
the true ground state of QCD with three
massless flavors at high density is the color-flavor locking
phase,  $K_{\gamma}=K$ for all $\gamma=u,d,s$. The condensate takes
\begin{equation}
\left<{\psi_L}^a_{i\alpha}(\vec v_F,x){\psi_L}^b_{j\beta}(-\vec v_F,x)
\right>=-\left<{\psi_R}^a_{i\alpha}(\vec v_F,x)
{\psi_R}^b_{j\beta}(-\vec v_F,x)\right>
=\epsilon_{ij}\epsilon^{abI}\epsilon_{\alpha\beta I}K(p_F),
\end{equation}
breaking the color symmetry, $U(1)_{\rm em}$, the chiral symmetry,
and the baryon number symmetry.
The symmetry breaking pattern of the CFL phase
is therefore
\begin{equation}
SU(3)_c\times SU(3)_L\times SU(3)_R\times U(1)_{\rm em}\times
U(1)_B\mapsto SU(3)_V\times U(1)_{\tilde Q}\times Z_2,
\end{equation}
where $SU(3)_V$ is the diagonal subgroup of three $SU(3)$ groups and
the generator of $U(1)_{\tilde Q}$ is a linear combination of the
color hypercharge and $U(1)_{\rm em}$ generator,
\begin{equation}
\tilde Q=\cos\theta Q_{\rm em}+\sin\theta Y_8,
\end{equation}
where $\tan\theta=e/g_s$.

Now, we analyze the SD gap equation to see if it admits a nontrivial
solution. Since the color-flavor locking gap is
preferred if it exists, we may write the gap as
\begin{equation}
\Delta_{\alpha\beta}^{ab}=\epsilon^{abI}\epsilon_{\alpha\beta I}\Delta.
\end{equation}
Then the gap equation becomes, neglecting small $h_i$ couplings,
\begin{equation}
\Delta(p_{\parallel})=(-ig_s)^2\int {{\rm d}^4l\over (2\pi)^4}
D_{\mu\nu}(p-l)V^{\mu}{T^a\Delta(l_{\parallel})(T^a)^T\over l_{\parallel}^2
-\Delta^2(l_{\parallel})}{\bar V}^{\nu}
+i{g_{\bar3}\over \mu^2}\int {{\rm d}^4l\over (2\pi)^4}
{\Delta(l_{\parallel})\over l_{\parallel}^2-\Delta^2(l_{\parallel})},
\label{gap}
\end{equation}
where we use the bare vertex and
take $Z(p_{\parallel})=1$ in the leading HDL approximation.

In the weak coupling limit, $|k_0|\ll|\vec k|$ and thus
\begin{eqnarray}
F(k_0,\vec k)\simeq M^2,\quad
G(k_0,\vec k)\simeq {\pi\over 4}M^2{k_0\over |\vec k|}.
\end{eqnarray}
Since the gap has to be fully antisymmetric in color indices, we get
\begin{eqnarray}
T_{tu}^a\Delta_{uv}(T^a)^T_{vs}=
\left({1\over 2}\delta_{tv}\delta_{us}-{1\over6}\delta_{tu}\delta_{vs}
\right)\Delta_{uv}=-{2\over 3}\Delta_{ts}
\end{eqnarray}
After Wick-rotating into Euclidean space, the gap equation becomes
\begin{eqnarray}
\Delta(p_{\parallel})\!&=&\!\int{d^4q\over (2\pi)^4}
\left[-{2\over3}g_s^2
\left\{{V\cdot P^T\cdot \bar V
\over (p-q)_{\parallel}^2+
\vec q_{\perp}^2+{\pi\over4}M^2|p_0-q_0|/
|\vec p-\vec q|}\right.\right.\nonumber\\
&-&\left.\left.{1\over (p-q)_{\parallel}^2+{\vec q_{\perp}}^2+M^2}
-\xi{(p-q)_{\parallel}^2\over
(p-q)^4}\right\}+{g_{\bar3}\over \mu^2}\right]
{\Delta(q_{\parallel})\over q_{\parallel}^2+\Delta^2(q_{\parallel})}.
\label{gap-hdl}
\end{eqnarray}
Note that the main contribution
to the integration comes from
the loop momenta in the region $q_{\parallel}^2\sim \Delta^2$ and
$|\vec q_{\perp}|\sim M^{2/3}\Delta^{1/3}$. Therefore, we find that
the leading contribution is by the first term due to the Landau-damped
magnetic gluons.
For this momentum range, we can take
$|\vec p-\vec q|\sim |\vec q_{\perp}|$ and
\begin{eqnarray}
V\cdot P^T\cdot{\bar V}=
-v_F^iv_F^j\left(\delta_{ij}-{\hat k}_i{\hat k}_j\right)
=-1+O\left({\Delta^{4/3}\over M^{4/3}}\right).
\end{eqnarray}
We also note that the term due to the four-Fermi operator is negligible,
since $g_{\bar3}\sim g_s^4$ at the matching scale $\mu$.

Neglecting $(p-q)_{\parallel}^2$
in the denominator, the gap equation becomes at the leading order
in the weak coupling expansion and $1/\mu$ expansion
\begin{eqnarray}
\Delta(p_{\parallel})={2g_s^2\over3}\int{d^4q\over (2\pi)^4}
\left[
{1\over \vec q_{\perp}^2+{\pi\over4}M^2|p_0-q_0|/|\vec q_{\perp}|}
+{1\over {\vec q_{\perp}}^2+M^2}
+\xi{(p-q)^2_{\parallel}\over |\vec q_{\perp}|^4}\right]
{\Delta(q_{\parallel})\over q_{\parallel}^2+\Delta^2(q_{\parallel})}.
\label{gap_leading}
\end{eqnarray}
The $\vec q_{\perp}$ integration can now be performed easily to get
\begin{eqnarray}
\Delta(p_{\parallel})={g_s^2\over 9\pi}
\int{d^2q_{\parallel}\over (2\pi)^2}{\Delta(q_{\parallel})\over
q_{\parallel}^2+\Delta^2}\left[\ln\left(
{\mu^3\over {\pi\over4}M^2|p_0-q_0|}\right)
+{3\over2}\ln\left({\mu^2\over M^2}\right)+{3\over2}\xi\right].
\end{eqnarray}
We see that in this approximation $\Delta(p_{\parallel})\simeq
\Delta(p_0)$. Then, we can integrate over $\vec v_F\cdot\vec q$
to get
\begin{eqnarray}
\Delta(p_0)&=& {g_s^2\over 36\pi^2}\int_{-\mu}^{\mu}dq_0
{\Delta(q_0)\over \sqrt{q_0^2+\Delta^2}}
\ln\left({{\bar\Lambda}\over |p_0-q_0|}
\right)
\label{gapf}
\end{eqnarray}
where $\bar\Lambda=4\mu/\pi\cdot (\mu/M)^5e^{3/2\xi}$.
If we take $\Delta\simeq \Delta(0)$ for a rough estimate of the gap,
\begin{equation}
1={g_s^2\over 36\pi^2}
\left[\ln\left({\bar\Lambda\over\Delta}\right)\right]^2
\quad
{\rm or}\quad
\Delta\simeq\bar\Lambda \exp\left(-{6\pi\over g_s}\right).
\end{equation}
As was done by Son~\cite{son}, one can convert the
Schwinger-Dyson gap equation (\ref{gapf})
into a differential equation to take into
account the energy dependence of the gap.
Approximating the logarithm in the gap equation as
\begin{equation}
\ln\left|p_0-q_0\right|\simeq
\left\{
\begin{array}{ll}
\ln\left|p_0\right| & \mbox{if $\left|p_0\right|>\left|q_0\right|$},\\
\ln\left|q_0\right| & \mbox{otherwise},
\end{array}
\right.
\end{equation}
we get
\begin{equation}
p\Delta^{\prime\prime}(p)+\Delta^{\prime}(p)+{2\alpha_s\over9\pi}
{\Delta(p)\over \sqrt{p^2+\Delta^2}}=0,
\label{diff}
\end{equation}
with boundary conditions $p\Delta^{\prime}=0$ at $p=\Delta$ and
$\Delta=0$ at $p=\bar\Delta$, where $p\equiv p_0$.
When $p\ll\Delta(p)$, the equation becomes
\begin{equation}
p\Delta^{\prime\prime}+\Delta^{\prime}+{r^2\over4}{\Delta(p)\over
|\Delta|}=0,
\end{equation}
where $r^2=2g_s^2/(9\pi^2)$ and $|\Delta|$ is the gap at $p=0$.
We find $\Delta(p)=|\Delta|J_0\left(r\sqrt{p/|\Delta|}\right)$
for $p\ll\|\Delta|$. When $p\gg \Delta$, the differential equation
(\ref{diff}) becomes
\begin{equation}
p\Delta^{\prime\prime}+\Delta^{\prime}+{r^2\over4}{\Delta\over p}=0,
\end{equation}
which has a power solution, $\Delta\sim p^{\pm ir/2}$.
Since the gap vanishes at $p=\bar\Lambda$, we get for $p\gg\Delta$
\begin{equation}
\Delta(p)=B\sin\left({r\over2}\ln{\bar\Lambda\over p}\right).
\end{equation}
By matching two solutions at the boundary $p=|\Delta|$ we get
\begin{equation}
B\simeq |\Delta|\quad {\rm and}\quad |\Delta|=\bar\Lambda
e^{-\pi/r}.
\end{equation}
The gap is therefore given as
at the leading order in the weak coupling expansion\footnote{
The gauge-parameter dependent term is subleading
in the gap equation (\ref{gap-hdl}).
Since the gap has to be gauge-independent, the gauge parameter
dependence in the prefactor will disappear if one includes
the higher order corrections.}
\begin{equation}
\left|\Delta\right|=
c\cdot{\mu\over g_s^5}\exp\left(-{3\pi^2\over\sqrt{2}g_s}\right),
\end{equation}
where $c=2^7\pi^4N_f^{-5/2}e^{3\xi/2+1}$. This agrees with
the RG analysis done by Son~\cite{son} (see also~\cite{hsu99})
and also with
the Schwinger-Dyson approach in full QCD~\cite{hmsw,sw99,pr99}.
The $1/g_s$ behavior of the exponent of the gap at high density is due to
the double logarithmic divergence in the gap equation~(\ref{gap-hdl}),
similarly to the case of chiral symmetry breaking under external magnetic
fields~\cite{hong96,miransky,miransky2}.
In addition to the usual logarithmic divergence
in the quark propagator as in the BCS superconductivity,
there is another logarithmic divergence due to the long-range gluon
exchange interaction, which occurs when the gluon loop momentum is colinear
to the incoming quark momentum ($\vec q_{\perp}\to0$).

\section{temperature effects and higher order corrections}

When the quark matter is not very dense and not very cold,
the effects of finite temperature and
density become important. In this section we calculate the
critical density and temperature.
First, we add the $1/\mu$ corrections to the gap equation
Eq.~(\ref{gap}) to see how the formation of
Cooper pair changes when the density decreases.
As derived in~\cite{hong98}, the leading
$1/\mu$ corrections to the quark-gluon interactions are
\begin{equation}
{\cal L}_1=-{1\over2\mu}\sum_{\vec v_F}\psi^{\dagger}(\vec
v_F,x)\left(\gamma_{\perp}\cdot D\right)^2\psi(\vec v_F,x)
=-\sum_{\vec v_F}\left[\psi^{\dagger}{D_{\perp}^2\over
2\mu}\psi+g_s\psi^{\dagger}{\sigma_{\mu\nu}F^{\mu\nu}\over
4\mu}\psi\right].
\end{equation}
In the leading order in the HDL approximation,
the loop correction to the vertex is
neglected and the quark-gluon vertex is shifted by the $1/\mu$
correction as
\begin{equation}
-ig_sv_F^i\mapsto -ig_sv_F^i-ig_s{l_{\perp}^i\over\mu},
\end{equation}
where $l_i$ is the momentum carried away from quarks by gluons.
Then, the gap equation (\ref{gap_leading}) becomes
\begin{eqnarray}
\Delta(p_{\parallel})&=&{2g_s^2\over 3}\int{{\rm d}^4l\over (2\pi)^4}
\left[{|\vec l_{\perp}|\left(1- l_{\perp}^2/\mu^2\right)\over
|\vec l_{\perp}|^3+(\pi/4) M^2|l_0-p_0|}
+{1-l_{\perp}^2/\mu^2\over l_{\perp}^2+M^2}
+\xi\cdot {l_{\parallel}^2\left
(1-l_{\perp}^2/\mu^2\right)\over l^4}\right]
{\Delta(l_{\parallel})\over l_{\parallel}^2+\Delta^2}.
\label{gap2}
\end{eqnarray}
For a constant gap approximation, $\Delta(p_{\parallel})
\simeq \Delta$, the gap equation becomes in the leading order,
as $p\to0$,
\begin{eqnarray}
1&=&{g_s^2\over9\pi}\int{{\rm d}^2l_{\parallel}\over
(2\pi)^2}\left[
\ln\left({\bar\Lambda\over |l_0|}\right)-{3\over2}\right]
{1\over l_{\parallel}^2+\Delta^2}\nonumber\\
&=&{g_s^2\over 36\pi^2}\ln\left({\bar\Lambda\over \Delta}\right)
\left[\ln\left({\bar\Lambda\over\Delta}\right)-3\right].
\end{eqnarray}
Therefore, we see that, when $\mu<\mu_c\simeq e^3\Delta$, the gap due
to the long-range color magnetic interaction disappears. Since the
phase transition for color superconducting phase is believed
to be of first order~\cite{phase,shuryak1}, we may assume that the gap
has the same dependence on the chemical potential $\mu$
as the leading order. Then, the critical density for the color
superconducting phase transition is given by
\begin{equation}
\mu_c=e^3\mu_c\exp\left[-{4\pi\sqrt{3}\over g_s(\mu_c)}\right].
\end{equation}
Therefore, if the strong interaction coupling is too strong
at the scale of the chemical potential, the gap does not form.
To form the Cooper pair gap, the strong coupling
at the scale of the chemical potential
has to be smaller than $g_s(\mu_c)=\pi^2/\sqrt{2}$.
By using the one-loop $\beta$ function
for three light flavors, $\beta(g_s)=-9/(16\pi^2)g_s^3$, and
the experimental value for the strong coupling constant,
$\alpha_s(1.73{\rm GeV})=0.32^{+0.031}_{-0.053}({\rm exp})
\pm0.016({\rm theo})$~\cite{ellis},
we get $0.13{\rm GeV}\lesssim\mu_c\lesssim0.31{\rm GeV}$, which
is about the same order as the one estimated by the instanton
induced four-Fermi interaction~\cite{shuryak1,shuryak2} or by general
effective four-Fermi interactions~\cite{phase}. But, this should
be taken as an order of magnitude,
since for such a small chemical
potential the higher order terms in $1/\mu$ expansion, which we
have neglected, are as important as the leading term.

So far we have not included the temperature effect in analyzing the
gap. The temperature effect is quite important to understand the
heavy ion collision or the final stage of the evolution of giant
stars because
quarks will have to have high energy to bring together to form a
super dense matter. The super dense and hot quark matter will go
through a phase transition as it cools down by emitting weakly
interacting particles like neutrinos.

At finite temperature,  $T$,  the gap equation (\ref{gap_leading})
becomes, following the imaginary-time formalism developed by
Matsubara~\cite{matsubara},
\begin{eqnarray}
\Delta(\omega_{n^{\prime}})={g_s^2\over 9\pi}T
\sum_{n=-\infty}^{+\infty}
\int{{\rm d}q\over 2\pi}{\Delta(\omega_n)\over \omega_n^2+
\Delta^2(\omega_n)+q^2}\ln\left({\bar\Lambda\over \left|
\omega_{n^{\prime}}-\omega_n\right|}\right),
\end{eqnarray}
where $\omega_n=\pi T(2n+1)$ and $q\equiv \vec v_F\cdot\vec q$.
We now use the constant
(but temperature-dependent) gap approximation,
$\Delta(\omega_n)\simeq \Delta(T)$ for all $n$.
Taking $n^{\prime}=0$ and converting the logarithm into integration,
we get
\begin{eqnarray}
\Delta(T)={g_s^2\over 18\pi}T\sum_{n=-\infty}^{+\infty}
\int{{\rm d}q\over 2\pi}\int_0^{{\bar\Lambda}^2}{\rm d}x
{\Delta(T)\over \omega_n^2+\Delta^2(T)+q^2}\cdot
{1\over x+(\omega_n-\omega_0)^2}.
\end{eqnarray}
Using the contour integral (see Fig.~4)~\cite{gusynin97},
one can in fact sum up over all $n$ to get
\begin{eqnarray}
1&=&{g_s^2\over36\pi^2}T\int {\rm d}q\int_0^{{\bar\Lambda}^2}
{\rm d}x{1\over 2\pi i}\oint_C{{\rm d}\omega\over 1+e^{-\omega/T}}
\cdot{1\over\left(\omega^2-q^2-\Delta^2(T)\right)\left[
(\omega_n-i\omega_0)^2+x\right]}.
\label{tgap}
\end{eqnarray}
Since the gap vanishes at the critical temperature,
$\Delta(T_C)=0$,
after performing the contour integration in Eq.~(\ref{tgap}),
we get
\begin{eqnarray}
1&=&{g_s^2\over36\pi^2}\int{\rm d}q\int_0^{{\bar\Lambda}^2}
{\rm d}x\left\{
{(\pi T_C)^2+x-q^2\over \left[
(\pi T_C)^2+x-q^2\right]^2+(2\pi T_Cq)^2}\cdot
{\tanh\left[q/(2T_C)\right]\over 2q}\right.\nonumber\\
& &\left.\quad\quad
+{(\pi T_C)^2+q^2-x\over \left[(\pi T_C)^2+q^2-x\right]^2
+(2\pi T_C)^2x}\cdot{ \coth\left[\sqrt{x}/(2T_C)\right]
\over \sqrt{2}}
\right\}.
\label{tgap1}
\end{eqnarray}
At high density $\bar\Lambda\gg T_C$,
the second term in the integral in Eq.~(\ref{tgap1}) is
negligible, compared to the first term, and
integrating over $x$, we get
\begin{eqnarray}
1&=&{g_s^2\over 36\pi^2}\int_0^{\lambda_c}{\rm d}y
{\tanh y\over y}\left[
\ln\left({\lambda_c^2\over (\pi/2)^2+y^2}\right)+
O\left({y^2\over \lambda_c^2}\right)\right]\nonumber\\
&=&{g_s^2\over 36\pi^2}\left[
\int_0^1{\rm d}y{\tanh y\over y}\ln\lambda_c^2
+\int_1^{\lambda_c}{\rm d}y{\tanh y\over y}\ln{\lambda_c^2\over y^2}
+\cdots\right]\\
&=&{g_s^2\over 36\pi^2}\left[
\left(\ln\lambda_c\right)^2+2A\ln\lambda_c+{\rm const.}\right]
\nonumber
\end{eqnarray}
where we have introduced $y\equiv q/(2T_C)$ and
$\lambda_c\equiv\bar\Lambda/(2T_C)$ and $A$ is given as
\begin{equation}
A=\int_0^1{\rm d}y{\tanh y\over y}+\int_1^{\infty}{\rm d}y
{\tanh y-1\over y}=\ln \left({4\over\pi}\right)+\gamma.
\end{equation}
Therefore, we find, taking the Euler-Mascheroni constant
$\gamma\simeq0.577$,
\begin{equation}
T_C={e^A\over2}\Delta\simeq1.13~\Delta.
\end{equation}
As comparison, we note in the BCS case, which has a contact
four-Fermi interaction with strength ${\bar g}$,
the critical temperature is given as
\begin{eqnarray}
1&=&{\bar g}\int_0^{\tilde\omega_c}{\rm d}z{\tanh z\over z}\nonumber\\
&\simeq&{\bar g}\left[\int_1^{\tilde\omega_c}{{\rm d}z\over z}
+\int_0^1{\rm d}z{\tanh z\over z}-\int_1^{\infty}{\rm d}z
{1-\tanh z\over z}\right]\\
&=&{\bar g}\ln\left(e^A\tilde\omega_c\right)\nonumber
\end{eqnarray}
where $\tilde\omega_c(\gg1)$ is determined by the Debye energy,
$\tilde\omega_c=\omega_D/(2T_C)$.
It shows that the ratio between the critical temperature and
the Cooper-pair gap\footnote{In the literature, the BCS gap is
defined as twice of the dynamical mass, $2\Delta$~\cite{schrieffer}.}
in color superconductivity is same as
the BCS value, $e^{\gamma}/\pi\simeq0.57$~\cite{pr99,brown99,talk99}.

\section{applications}
\label{sec:label7}

It is believed that the core of compact stars like neutron stars
may be dense enough to form quark matter and may shed
some lights on understanding QCD at high density.
The properties of compact stars can be investigated by studying
the emission of weakly interacting particles like neutrinos
or axions, which is the dominant
cooling process of the compact stars~\cite{raffelt,hong98ns}.
Since the emission rate depends on the couplings of those particles,
it is important to understand how the interaction and the coupling of
neutrinos or axions change in dense quark matter.

Neutrinos interact with quarks by the exchange of neutral
currents, which is described, at low energy, as four-Fermi
interaction,
\begin{equation}
{\cal L}_{\nu q}={G_F\over\sqrt{2}} \bar \Psi_L\gamma^{\mu}
\Psi_L\bar \nu_L\gamma_{\mu}\nu_L,
\end{equation}
where $G_F=1.166\times10^{-5}~{\rm GeV}^{-2}$ is the Fermi
constant.
Again, by decomposing the quark fields as in Eq.~(\ref{decomp})
and integrating out $\psi_-$ modes, the four-Fermi interaction
becomes
\begin{equation}
{\cal L}_{\nu q}={G_F\over\sqrt{2}}\sum_{\vec v_F}
{\psi_{+}^{\dagger}}_{L}(\vec v_F,x){\psi_{+}}_{L}
(\vec v_F,x) \bar\nu_L(x)\!
\mathrel{\mathop{V\!\!\!\!/}}\nu_L(x)+\cdots,
\end{equation}
where the ellipsis denotes the higher order terms in the power
expansion of $1/\mu$.
Since the four-fermion interaction of quarks with opposite momenta
gets enhanced a lot at low energy, as
we have seen in the previous section (Sec.~\ref{sec:label6}),
it may have significant corrections to the couplings of those
weakly interacting particles to quarks.
We first calculate the one-loop correction to the neutrino-quark
four-Fermi coupling by the marginal four-quark interaction:
\begin{eqnarray}
\delta{\cal L}_{\nu q}&=&{G_F\over\sqrt{2}}
{\psi_{+}^{\dagger}}_{L}(\vec v_F,x){\psi_{+}}_{L}
(\vec v_F,x) \bar\nu_L(x)\!
\mathrel{\mathop{V\!\!\!\!/}}\nu_L(x)\nonumber\\
& &\times {ig_{\bar3}\over 2 M^2}\delta^A_{tv;us}\int_y
\left[\bar \psi_t(\vec v_F^{\prime},y)\gamma^0
\psi_s(\vec v_F^{\prime},y)
\bar\psi_v(-\vec v_F^{\prime},y)\gamma^0\psi_u(-\vec v_F^{\prime},y)
\right]\\
&=&{4\over 3}{g_{\bar 3}\over 2\pi}{G_F\over\sqrt{2}}
{\psi_{+}^{\dagger}}_{L}(\vec v_F,x){\psi_{+}}_{L}
(\vec v_F,x) \bar\nu_L(x)\bar V\cdot\gamma\nu_L(x),
\nonumber
\end{eqnarray}
where $\vec v_F$ and $\vec v_F^{\prime}$ are summed over and $g_{\bar3}$
is the value of the marginal four-quark coupling at the screening mass
scale $M$.
Now, in fact, because of the kinematical constraint due to the
presence of the Fermi surface, only the cactus diagrams,
shown in Fig.~5, contribute
to the coupling corrections, which can be summed up as
\begin{eqnarray}
\delta {\cal L}_{\nu q}&=&{4G_F\over3\sqrt{2}}
\sum_{\vec v_F}
\left[\left({g_{\bar3}\over 2\pi-g_{\bar3}}\right)
{\psi_{+}^{\dagger}}_{L}(\vec v_F,x){\psi_{+}}_{L}
(\vec v_F,x) \nu_L^{\dagger}(x)\nu_L(x)\right.\\
& &\left.
-\left({g_{\bar3}\over 2\pi+g_{\bar3}}\right)
{\psi_{+}^{\dagger}}_{L}(\vec v_F,x){\psi_{+}}_{L}
(\vec v_F,x) \bar\nu_L(x)\vec v_F\cdot \vec \gamma\nu_L(x)
\right].
\nonumber
\end{eqnarray}
Similarly, for axions which couple to quarks as
\begin{equation}
{\cal L}_{aq}={1\over 2f_{PQ}}\partial_{\mu}a\bar
\Psi\gamma^{\mu}\gamma_5\Psi,
\end{equation}
where $a$ is the axion field and $f_{PQ}$
is the axion decay constant,
we find the correction to the axion-quark coupling to be
\begin{eqnarray}
\delta {\cal L}_{aq}&=&{2\over 3f_{PQ}}
\sum_{\vec v_F}
\left[\left({g_{\bar3}\partial_0a\over 2\pi-g_{\bar3}}\right)
{\psi_{+}^{\dagger}}(\vec v_F,x)\gamma_5{\psi_{+}}
(\vec v_F,x)
-\left({g_{\bar3}\vec v_F\cdot\vec \nabla a\over 2\pi+g_{\bar3}}
\right){\psi_{+}^{\dagger}}(\vec v_F,x)\gamma_5{\psi_{+}}
(\vec v_F,x)
\right].
\nonumber
\end{eqnarray}

Therefore, as the marginal four-quark coupling approaches $2\pi$,
the quark-neutrino and quark-axion couplings become divergent.
Since at low energy $g_{\bar3}$ is quite large in dense quark
matter, we argue that
quark matter will produce neutrinos or axions copiously if
the density of quark matter is high enough ($\mu\gg\Lambda_{\rm QCD}$)
such that the marginal four-quark interaction gets enhanced
sufficiently.



\section{conclusion}
\label{sec:label8}
We have studied in detail an effective field theory of QCD at high
density, constructed by integrating out the anti-quarks
to describe the low energy dynamics of dense quark matter.
In the effective theory, the dynamics of quarks is
effectively (1+1)-dimensional; the energy of quarks does not
depend on the perpendicular momentum to the Fermi velocity,
which just serves to label the degeneracy.
At energy lower than the screening mass of electric gluons,
the relevant interactions are four-Fermi interactions with
opposite incoming momenta and the coupling with magnetic gluons.

Because of the dimensional reduction at high density,
both four-Fermi interaction and magnetic gluon exchange
interaction lead to Cooper-pair condensate of quarks in color
anti-triplet channel for arbitrarily weak coupling. In the ladder
approximation, the gap formed by the magnetic gluon exchange
interaction is much bigger than the one by four-Fermi interaction.
The color-flavor locking condensate is found to be energetically
more preferred for high density quark matter
with three light flavors, because the
color-flavor locking gap is bigger than the spin one gap,
if we include the four-Fermi interactions.
Further, including the $1/\mu$ corrections and the temperature
effects, the critical density and the
critical temperature for color superconducting phase are calculated
by solving the gap equation in the HDL approximation.

Finally, we have calculated the corrections to the quark-neutrino
four-Fermi interaction and to the quark-axion coupling in dense
quark matter due to the marginal four-quark interaction.
And we found the correction is quite significant and thus
dense quark matter will copiously produce neutrinos and
axions.

\acknowledgments

The author is grateful to Sekhar Chivukula, Andy Cohen, Raman
Sundrum for useful comments, and wishes to thank Roman Jackiw,
Steve Hsu,
Cristina Manuel, Mannque Rho, Dirk Rischke, Tom Sch\"afer, and
Andrei Smilga for enlightening discussions.
The author wishes to acknowledge the financial support of the Korea
Research Foundation made in the program year of 1998
(1998-15-D00022). This work was also supported by Pusan National
University Research grant, 1998.

\newpage
\begin{figure}
\vskip 0.2in

\centerline{\epsfbox{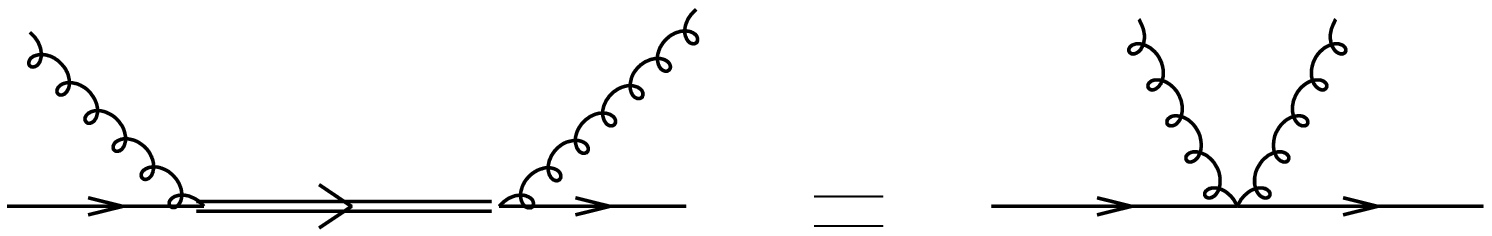}}
\vskip 0.2in
\caption{The tree-level matching condition. Wiggly lines denote gluons;
solid lines, states near the Fermi surface;
and double solid line, states in the Dirac sea. }
\end{figure}
\vskip 0.2in
\begin{figure}
\centerline{\epsfbox{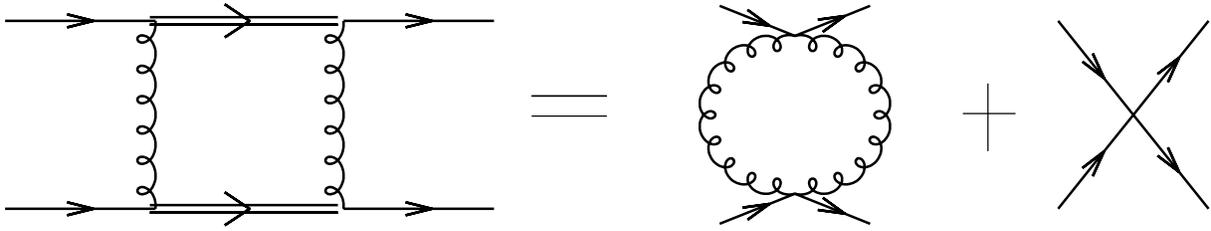}}
\vskip 0.2in
\caption{The one-loop matching condition for a four-quark amplitude.}
\end{figure}
\vskip 0.2in

\begin{figure}
\centerline{\epsfbox{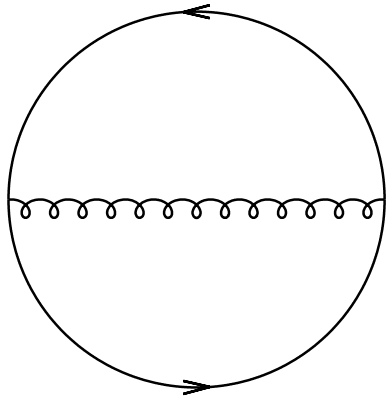}}
\vskip 0.2in
\caption{The 2PI diagram in the leading order HDL approximation.}
\end{figure}
\begin{figure}
\centerline{\epsfbox{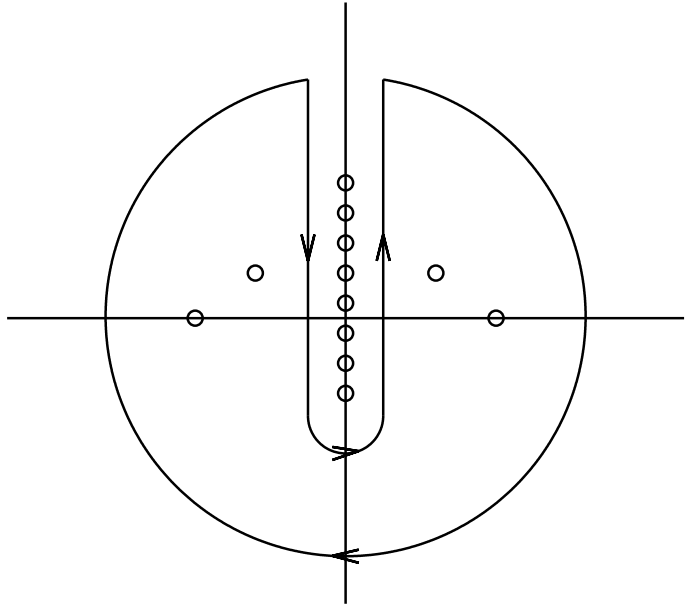}}
\vskip 0.2in
\caption{The contour for integration over the Matsubara frequency.
The circles are poles.}
\end{figure}

\begin{figure}
\centerline{\epsfbox{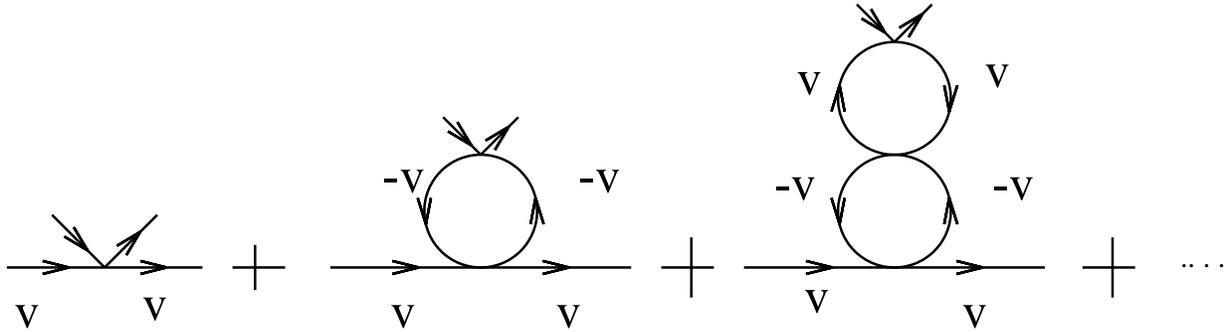}}
\vskip 0.2in
\caption{The corrections to the quark-neutrino four-Fermi coupling.}
\end{figure}

\end{document}